\documentclass[prl,twocolumn,nofootinbib,superscriptaddress, floatfix,nobibnotes]{revtex4-2}

\usepackage{xcolor}
\usepackage{graphicx}
\usepackage{amsmath,amsthm,amsfonts,amssymb,mathtools}

\usepackage{siunitx}                                            % numbers with units
\usepackage{hyperref}
\hypersetup{hidelinks}

\usepackage{microtype}

\makeatletter
\renewcommand{\frontmatter@finalspace}{\addvspace{8pt}}
\makeatother

\AtBeginDocument{%
    \setlength{\skip\footins}{24pt plus 2pt minus 2pt}%
    \renewcommand{\footnoterule}{%
        \kern-10pt
        \hrule width .5in
        \kern10pt
    }%
}

\begin{document}
\title{Non-equilibrium orientational dynamics of pinned topological defects\texorpdfstring{\\}{ }
in two-dimensional nematic liquid crystals}

\author{Petur Bryde}
\altaffiliation[Present address: ]{University of Iceland, Reykjavík, Iceland}

\author{J\"orn Dunkel}
\affiliation{Department of Mathematics, Massachusetts Institute of Technology, Cambridge, MA, USA}

\date{November 2, 2026}

\begin{abstract}
    The long-wavelength dynamics of two-dimensional (2D) nematic liquid
    crystals (LCs) is dominated by the motion of topological defects, which
    behave like interacting particles endowed with both topological charge and
    orientation. In experimentally relevant regimes, the translational dynamics
    of the defect positions can be accurately described by effective
    nematoelastic particle-particle interaction forces. By contrast, analogous
    quasistatic torque descriptions for the orientational dynamics typically
    fail due to a breakdown of time-scale separation, severely hindering the
    inverse design of texture response dynamics in nematic LCs. To overcome
    this obstruction, we introduce a linear phase-diffusion formulation that
    accurately predicts the orientational relaxation dynamics of pinned
    topological defects in the nonlinear 2D Landau--de Gennes theory.
    The broad practical potential of this approach is demonstrated by showing
    how it enables the inverse design of transient nematic patterns as well as
    the inference of hidden defect configurations from sparse measurements.
\end{abstract}
\maketitle
\widowpenalties 3 10000 10000 0\relax

Topological defects in ordered media can act as robust, controllable degrees of
freedom for transmitting signals~\cite{zhang2022logic},
storing information~\cite{meng2025emergent,kos2022nematic},
and programming material response~\cite{shankar2024design}.
In liquid crystals, defects can be created, pinned, and reconfigured using
colloidal inclusions~\cite{musevic2006two,machon2013knots}, confinement and boundary
design~\cite{tkalec2011reconfigurable,campos2026shaping,fernandez2007novel,
guo2021photopatterned,wu2025topological},
electric fields~\cite{sandford2020electrically}, and optical and optothermal
forcing~\cite{emervsivc2019sculpting}.
In active nematics~\cite{doostmohammadi2018active},
light-patterned activity provides an additional route for creating and controlling
defects~\cite{zhang2021spatiotemporal}.
The large-scale dynamics of nematic systems can often be reduced to the motion
of topological defects, which behave as quasiparticles whose interactions are
mediated by the surrounding director field~\cite{bowick2009two,turner2010vortices}.
This defect-level description is attractive for control applications because it
recasts continuum-field optimization in terms of a few collective
coordinates~\cite{zhang2022logic,shankar2024design}.

% Figure 1
\begin{figure*}[!t]
    \includegraphics[scale=0.9]{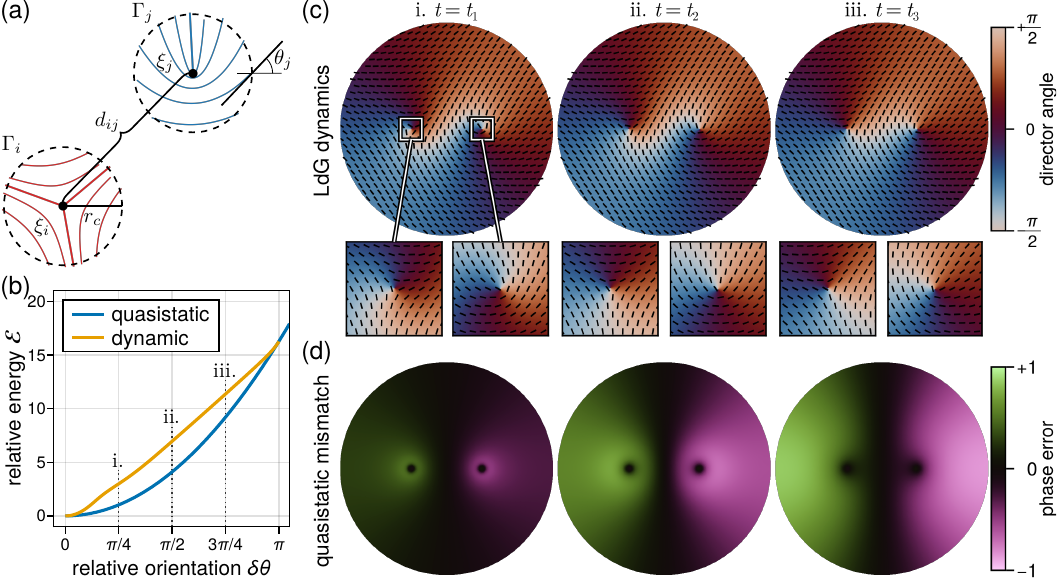}
    \caption{
        Breakdown of the quasistatic approximation for orientational defect dynamics due to
        lack of time-scale separation.
        (a) Schematic of the pinned two-defect setup. The defects are centered at fixed
        positions \(\xi_i\) with separation \(d_{ij}=|\xi_i-\xi_j|\). Their
        orientations \(\theta_i\) are determined from the local nematic texture within
        circles \(\Gamma_{i}\) of radius \(r_c\), the characteristic defect core radius.
        The relative orientation \(\delta\theta=\theta_2-\theta_1\) parameterizes a family
        of twisted quasistatic textures obtained by minimizing the energy subject to fixed
        defect orientations.
        (b) The Landau--de Gennes (LdG) relaxation, initialized from a twisted quasistatic
        texture, quickly leaves this manifold of constrained minimizers. The dynamic energy
        trajectory (orange) lies above the quadratic quasistatic energy \(E_{\mathrm{qs}}\)
        evaluated at the same instantaneous \(\delta\theta\), showing that the defect
        orientations alone do not determine the evolving nematic texture.
        Both energies are measured relative to the energy of the equilibrium state \(Q_0\).
        Labels i--iii mark the times shown in panels (c) and (d).
        (c) Snapshots of the evolving nematic texture show bulk relaxation while the pinned
        defects rotate. Insets show an enlarged view of the defect near-field. (d) The phase
        difference between the evolving nematic texture and the quasistatic texture with the
        same \(\delta\theta\) initially develops near the defects and then spreads through
        the bulk, showing that the evolving texture is not represented by the corresponding
        constrained minimizer.
        The domain is a disk of radius \(150\), and the defect separation is \(100\), both
        in units of \(\ell/\sqrt{2}\).
    }
    \label{fig:quasistatic-breakdown}
\end{figure*}

Besides position and topological charge~\cite{alexander2012colloquium},
nematic defects possess orientational degrees of freedom that provide additional
collective coordinates~\cite{vromans2016orientational,tang2017orientation,tang2019theory}.
These orientations affect the elastic interactions between defects, leading to coupled
translational and orientational dynamics~\cite{vromans2016orientational,pearce2021properties,pearce2021orientational}.
In active nematics, they also govern the self-propulsion of \(+1/2\)
defects~\cite{giomi2013defect,pismen2013dynamics,shankar2022topological}.
Defect orientations have also been proposed as information-bearing variables
in nematic bits~\cite{kos2022nematic,lei2026braiding}.

Reflecting the practical importance of defect control, substantial progress
has been made in developing reduced models of defect translation in both
passive~\cite{lin1998complex,jerrard1998dynamics,sandier2004gamma} and
active~\cite{giomi2013defect,shankar2018defect,vafa2020multidefect}
nematics.
By contrast, no corresponding effective theory has been established for the
\emph{orientational} dynamics of defects.
This gap currently prevents the design of efficient control strategies
for systems of nematic bits, as approaches based on the full nematic field equations require
solving computationally expensive PDE-constrained optimization
problems~\cite{surowiec2023optimal,norton2020optimal}.
Here, we develop a linear reduced model that captures the~\mbox{coupled} relaxation
of defect orientations and the surrounding nematic texture, providing
a tractable framework for inverse design and control.

For the translational dynamics of defects, the usual quasistatic
reduction~\cite{lin1998complex,jerrard1998dynamics,sandier2004gamma}
relies on a time-scale separation between defect motion and the relaxation of the
surrounding director field. This allows the bulk texture to remain close to the
energy-minimizing solution determined by the instantaneous defect positions, thereby
yielding effective elastic forces~\cite{imura1973interaction} and finite-dimensional
equations of motion for the defect centers~\cite{pismen1992motion,denniston1996disclination}.
Pioneering previous work~\cite{keber2014topology,vromans2016orientational} has applied the
same quasistatic approach to defect orientations, treating them as prescribed defect data
and deriving orientation-dependent energies and torques in
two-defect~\cite{tang2017orientation} and many-defect
configurations~\cite{pearce2021properties,vcopar2024many}.
However, while the quasistatic reduction succeeds for defect translation, it fails to capture the
orientational dynamics of defects.
We demonstrate this by studying purely relaxational
dynamics of pinned defects in the two-dimensional Landau--de Gennes (LdG) model. This
breakdown arises from the absence of time-scale separation between defect orientations and
the surrounding director field. Consequently, the evolving nematic texture is not determined
by the defect orientations alone.

Here, we overcome this limitation by formulating the orientational dynamics in terms of a
scalar relative phase field measured with respect to a reference equilibrium texture and
governed by a linear diffusion equation. This diffusing bulk degree of freedom
captures the rotations of pinned defects while treating \(+1/2\) and \(-1/2\) charges in a
unified way.
Importantly from a practical perspective, the reduction turns nonlinear PDE-constrained
optimization problems for the LdG dynamics into more tractable problems with a linear state
equation, marking a step toward design of defect-based programmable
materials~\cite{zhang2022logic,shankar2024design}. We demonstrate the broad
practical potential of this reduction through two archetypal applications: inverse design of
transient nematic patterns and recovery of hidden initial defect configurations from sparse
measurements.

\widowpenalties 0\relax

To isolate orientational relaxation, we study pinned defects in purely
relaxational LdG dynamics, neglecting hydrodynamic backflow~\cite{toth2002hydrodynamics}.
In the 2D setting considered here, the symmetric
traceless tensor order parameter has two independent components and can be
represented by a complex order parameter \(Q\)~\cite{majumdar2012landau}, normalized so that
\(|Q| = 1\) in the uniform ordered state.
We work in dimensionless variables, rescaling lengths by the nematic coherence length
\(\ell\) and time by the corresponding diffusive relaxation time \(\ell^2/D\), where \(D\)
is the orientational diffusivity.
The dimensionless LdG free energy on a domain \(\Omega\subset\mathbb C\)
is taken as~\cite{deGennes1993physics,chaikin1995principles}
\begin{equation}
    \label{eq:free-energy}
    \mathcal{F}(Q) = \int_\Omega
    \left[
        \frac{1}{2} |\nabla Q|^2 + \frac{1}{4} (\alpha-|Q|^2)^2
    \right] \mathrm{d}^{2}x.
\end{equation}
The spatially varying coefficient \(\alpha\) pins defects at prescribed positions
\(\xi_j\in\Omega\) (\(j=1,\ldots,N\)) by locally suppressing the preferred scalar order
parameter \(s=|Q|\): \(\alpha\) is negative at the pinning sites and increases smoothly to
\(\alpha=1\) over a distance of order unity in the rescaled coordinates.
Functionals of the type~\eqref{eq:free-energy} have previously been studied as a model of
vortex pinning by impurities in superconductors~\cite{rubinstein1995equilibrium}.
With the defect positions fixed by the pinning field, we consider the relaxation
of their orientations and the surrounding nematic texture, governed by
\begin{equation}
    \label{eq:ldg-relaxation}
    \partial_t Q = \Delta Q + \left(\alpha-|Q|^2\right)Q.
\end{equation}
For simplicity, we assume free (Neumann) boundary conditions on \(\partial \Omega\)
throughout.

% Figure 2
\begin{figure*}[!t]
\includegraphics[scale=0.9]{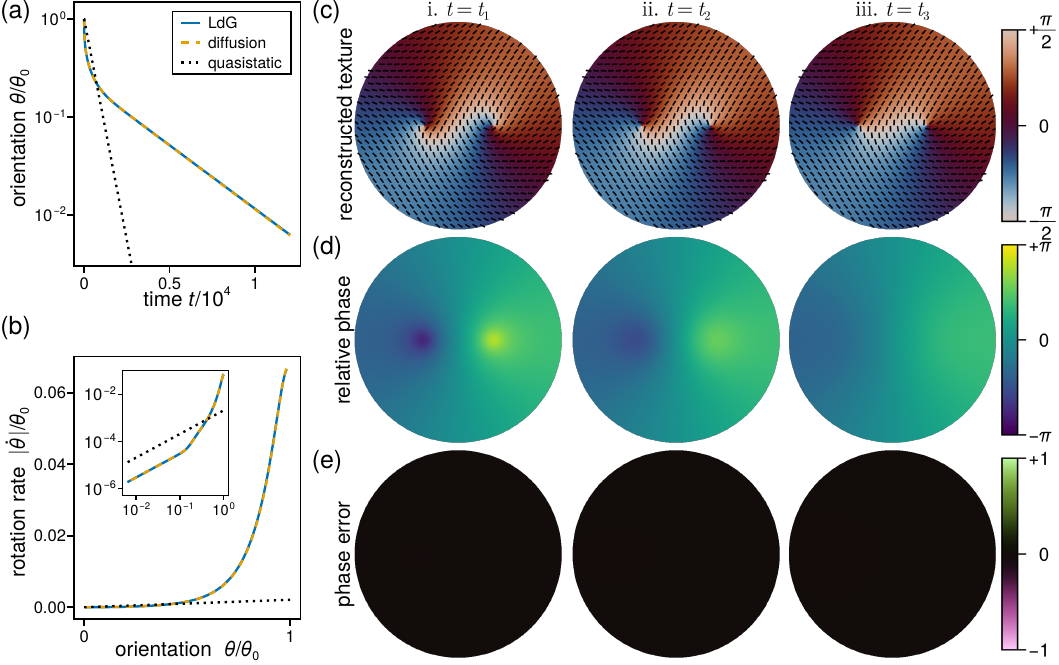}
\caption{
    Linear phase diffusion quantitatively captures two-defect orientational relaxation.
    (a) Relaxation of the defect orientation \(\theta(t)\), normalized
    by its initial value \(\theta_0\). Landau--de Gennes (LdG) curves for
    different \(\theta_0\) collapse under this normalization up to small
    early-time deviations; the representative solid curve shown has
    \(\theta_0=\pi/2\). The relative phase diffusion model (dashed line)
    follows the collapsed LdG relaxation. The dotted quasistatic prediction
    follows \(\dot{\theta}=-\kappa\theta\), giving exponential decay with
    quasistatic decay rate \(\kappa\).
    Time is measured in units of the diffusive relaxation time \(\ell^2/D\).
    (b) Plotting the rotation rate against the orientation, both normalized by the initial
    orientation \(\theta_0\), directly compares the dynamical laws. The quasistatic torque
    law predicts a straight line, whereas the LdG dynamics follow a nonlinear
    relation captured by the relative phase diffusion model.
    (c) Reconstructing the nematic texture from the diffusing relative phase
    reproduces the LdG result at representative times during relaxation.
    (d) The relative phase \(\varphi=\arg(Q/Q_0)\) between the
    complex order parameter \(Q\) and a reference equilibrium state \(Q_0\)
    measures the local rotation away from equilibrium; in this representation, the
    relaxation appears as diffusion of a smooth phase field.
    (e) Pointwise phase error relative to the full Landau--de Gennes
    relaxation, shown on the same scale as Fig.~\ref{fig:quasistatic-breakdown}(d); the maximum absolute
    error over the displayed snapshots is below \(0.06\), less than \(2\%\) of
    the initial phase amplitude \(\pi\).
    The domain and defect positions are the same as in Fig.~\ref{fig:quasistatic-breakdown}.
}
\label{fig:fig2}
\end{figure*}

For the prescribed configuration of pinned defects, we denote the
corresponding equilibrium texture by \(Q_0\), which, away from the
defect cores, is well approximated by
\begin{equation}
    \label{eq:reference-texture}
    Q_0(x) \simeq s_0(x)
    \exp\Big[
        2i\Big(
            \phi_0 +
            \sum_{j=1}^{N} k_j \arg(x-\xi_j)
        \Big)
    \Big],
\end{equation}
where \(k_j = \pm 1/2\) are the topological charges and \(\phi_0\) is an arbitrary
global phase.
Near each defect center, the scalar order parameter satisfies
\(s_0(x)\sim |x-\xi_j|/r_c\), which defines the characteristic
defect-core radius \(r_c\), while \(s_0\approx 1\) away from the cores.
We describe deviations from this equilibrium by introducing the relative phase
\(\varphi\), defined as a continuous branch of \(\arg(Q/Q_0)\).
The singular winding of \(Q\) and \(Q_0\) cancels in
this ratio, so unlike the director angle, \(\varphi\) is single-valued and
extends continuously into the defect cores.
Similarly, we measure energies relative to the reference equilibrium,
defining \(\mathcal{E}(Q)=\mathcal{F}(Q)-\mathcal{F}(Q_0)\).

We next define the defect orientations used to compare the LdG relaxation with the
quasistatic reduction. In this setting, only relative defect
orientations affect the orientational dynamics, so we define each orientation as
a deviation from the reference equilibrium \(Q_0\).
This is conveniently expressed in terms of the relative phase by setting
\(\theta_j=\frac{1}{2}\langle\varphi\rangle_{j}\), where \(\langle\varphi\rangle_{j}\)
denotes the average over the disk of radius \(r_c\) centered at the defect position
\(\xi_j\), whose boundary we denote \(\Gamma_j\) (Fig.~\ref{fig:quasistatic-breakdown}(a)).
The factor \(1/2\) appears because the relative phase
\(\varphi\) is twice the local director
rotation relative to \(Q_0\). This definition treats \(+1/2\) and \(-1/2\)
defects symmetrically and gives \(\theta_j=0\) for all \(j\) in the reference
texture.
In the quasistatic approximation, the instantaneous nematic state is replaced by the
quasistatic texture \(Q_{\mathrm{qs}}(x;\boldsymbol{\theta})\), defined as the minimizer
of \(\mathcal{F}\) subject to prescribed defect orientations
\(\boldsymbol{\theta}=(\theta_1,\ldots,\theta_N)\).
Away from the defect cores, its relative phase is approximately harmonic
and admits the logarithmic representation
\begin{equation}
    \label{eq:spiral-charge-relative-phase}
    \varphi_{\mathrm{qs}}(x;\boldsymbol{\theta})
    \simeq 2 \sum_{j=1}^{N} \mu_{j} \log \frac{|x-\xi_j|}{r_c},
\end{equation}
where the coefficients \(\boldsymbol{\mu}=(\mu_1,\ldots,\mu_N)\),
introduced by Čopar and Kos~\cite{vcopar2024many} as spiral
charges, are linearly related to \(\boldsymbol{\theta}\).

Restricting the relative energy to the family of quasistatic solutions defines
a constrained energy landscape
\(E_{\mathrm{qs}}(\boldsymbol{\theta}) = \mathcal{E}(Q_{\mathrm{qs}}(\cdot;\boldsymbol{\theta}))\),
whose gradients determine the elastic torques on the defect orientations.
Orientational interaction energies of this type have been explicitly calculated for
two defects by Tang and Selinger~\cite{tang2017orientation} and
extended to many-defect configurations by Čopar and Kos~\cite{vcopar2024many}.
By analogy with quasistatic reductions for defect translation, this suggests
replacing the full LdG relaxation by a constrained gradient flow on this landscape,
\(\mathbf{G}\dot{\boldsymbol{\theta}}=-\nabla E_{\mathrm{qs}}(\boldsymbol{\theta})\),
where \(\mathbf{G}\) is the induced metric, or inverse
mobility, on the manifold of twisted quasistatic textures.
Using the spiral charge representation~\eqref{eq:spiral-charge-relative-phase}
for the quasistatic texture, and neglecting the weak dependence of the
scalar order parameter on defect orientations, one finds that
\(E_{\mathrm{qs}}\) is quadratic in \(\boldsymbol{\theta}\) and that \(\mathbf{G}\)
is constant. Consequently, the constrained gradient flow predicts restoring
torques that depend linearly on \(\boldsymbol{\theta}\), leading to exponential relaxation.

We test the quasistatic assumption in the minimal two-defect geometry, with
pinned defect centers separated by \(d_{ij}=|\xi_i-\xi_j|\) and orientations
described by the relative orientation \(\delta\theta=\theta_2-\theta_1\)
(Fig.~\ref{fig:quasistatic-breakdown}(a)).
Starting from a twisted quasistatic texture with nonzero \(\delta\theta\), we
evolve the full LdG dynamics while tracking both \(\boldsymbol{\theta}(t)\) and
the relative energy \(E(t) = \mathcal{E}(Q(t))\). If the quasistatic reduction were valid,
the evolving state would remain close to the constrained minimizer with the same
instantaneous defect orientations. Plotting \(E(t)\) against \(\delta \theta(t)\) would
then reproduce the quasistatic energy curve \(E_{\mathrm{qs}}\).
Instead, we find that the relative energy \(E(t)\) lies above the quasistatic energy
soon after initialization, showing that the nematic texture leaves the manifold of twisted
quasistatic solutions (Fig.~\ref{fig:quasistatic-breakdown}(b)). Snapshots of the nematic
texture and the phase difference relative to the corresponding quasistatic texture show that
the deviations initially develop near the defects and then spread through the bulk
(Fig.~\ref{fig:quasistatic-breakdown}(c),(d)).

The failure of the quasistatic approximation shown in
Fig.~\ref{fig:quasistatic-breakdown} reflects the absence of
time-scale separation between defect orientations and the surrounding texture.
Consequently, the relative phase \(\varphi\) cannot be assumed to retain
the quasistatic form in Eq.~\eqref{eq:spiral-charge-relative-phase} under LdG relaxation.
We therefore treat the relative phase as an independent bulk dynamical field.
Approximating the scalar order parameter by its equilibrium profile \(s \approx s_0 =
|Q_0|\), we have \(Q(x,t)\approx Q_0(x)e^{i\varphi(x,t)}\). Substituting into
Eq.~\eqref{eq:ldg-relaxation} and taking the phase component yields the reduced equation
\begin{equation}
    \label{eq:phase-diffusion}
    s_0^2 \partial_t \varphi = \nabla \cdot \left(s_0^2\nabla\varphi\right).
\end{equation}
Away from defect cores, \(s_0\approx 1\), so Eq.~\eqref{eq:phase-diffusion}
approximately reduces to the scalar heat equation. In this representation, the
defect orientations are not autonomous coordinates; they are determined by the
relative phase, which relaxes through the bulk.

For the symmetric two-defect relaxation considered above, the phase-diffusion model
reproduces the evolution of the defect orientation \(\theta=\theta_2=-\theta_1\) obtained
from the full LdG dynamics. The quasistatic gradient flow instead predicts
\(\dot{\theta}=-\kappa\theta\) with constant decay rate \(\kappa\)
and fails to capture both the transient dynamics and long-time decay rate
(Fig.~\ref{fig:fig2}(a)).
The same comparison can be made in the phase plane: plotting the
rotation rate \(\dot{\theta}\) against \(\theta\) shows that linear phase diffusion
reproduces the nonlinear relation observed in the LdG evolution,
whereas the quasistatic theory predicts an effective torque linear in the orientation
(Fig.~\ref{fig:fig2}(b)).
The relative phase also allows us to reconstruct the evolving nematic
texture with small pointwise phase errors relative to the full LdG
solution (Fig.~\ref{fig:fig2}(c)--(e)).

% Figure 3
\begin{figure}[!t]
\includegraphics[scale=1]{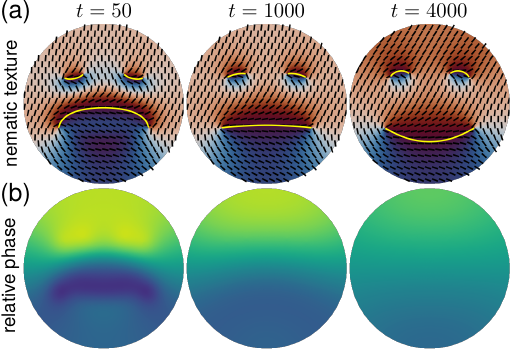}
\caption{
    Linear phase diffusion captures many-defect texture evolution. A designed
    initial relative phase is chosen to produce a frown-like pattern in the nematic
    texture, which evolves through a smile-like transient state. Columns show
    snapshots at times \(t=50\), \(1000\), and \(4000\) in units of the diffusive
    relaxation time \(\ell^2/D\). (a) Nematic texture reconstructed from the diffusion
    model, with yellow curves marking \(\arg Q=0 \pmod{2\pi}\), where the director is
    parallel to the \(x\)-axis.
    (b) The same evolution in the relative phase representation
    \(\varphi=\arg(Q/Q_0)\). Localized perturbations in the early-time state
    diffuse rapidly, leaving a slowly relaxing long-wavelength mode. Color scales
    match those in Fig.~\ref{fig:fig2}(c)--(d).
    The pointwise phase error against the full Landau--de Gennes relaxation remains below
    \(0.07\) over the displayed snapshots, less than \(2\%\) of the initial phase amplitude.
    The domain is a disk of radius \(200 \ell/\sqrt{2}\).
}
\label{fig:fig3}
\end{figure}

% Figure 4
\begin{figure*}[!t]
\includegraphics[scale=0.9]{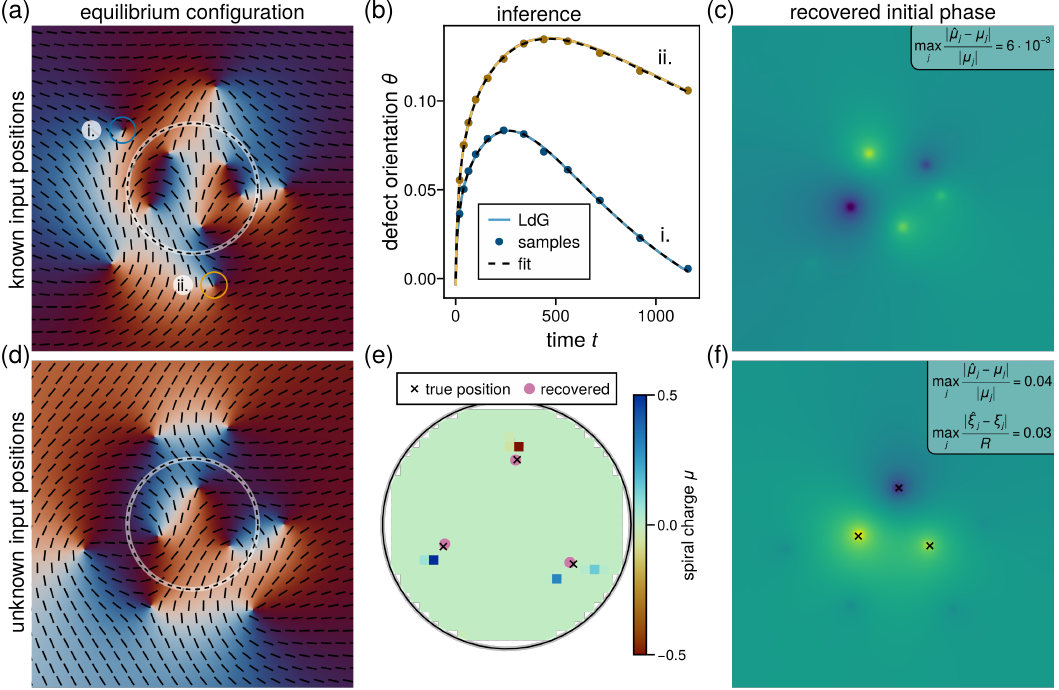}
\caption{
    Linear phase diffusion recovers hidden defect orientations and positions from
    sparse measurements. Input defects inside a central disk of radius \(R\)
    have unknown initial orientations relative to a reference equilibrium
    texture, while the orientations of output defects outside the disk are
    measured during relaxation. The number and positions of inputs are known
    (top row) or unknown (bottom row).
    Noisy synthetic measurements are generated from nonlinear Landau--de Gennes
    (LdG) relaxation, while inference uses the linear relative phase diffusion model.
    (a) Reference equilibrium for the known-position case with five inputs and
    five outputs. Two representative measured outputs, labeled i and ii, are
    highlighted.
    (b) Output orientations \(\theta(t)\) for defects i and ii in panel (a).
    Time is measured in units of the diffusive relaxation time \(\ell^2/D\).
    Solid curves show noiseless LdG output orientations; noisy
    measurements are shown as points.
    Dashed curves show linear least-squares fits using all output measurements
    at the sampled times.
    (c) Recovered initial relative phase \(\varphi(x,0)\) reconstructed from
    the inferred spiral charges \(\hat{\mu}_j\).
    (d) Unknown-position case with three hidden inputs and six measured
    outputs.
    (e) Sparse recovery on a grid of candidate input positions.
    Colors show the recovered sparse vector of spiral-charge coefficients on the candidate
    grid, circles mark the inferred defect positions \(\hat{\xi}_j\) after off-grid
    refinement, and black crosses mark the true input positions.
    (f) Recovered initial relative phase \(\varphi(x,0)\) reconstructed
    from the inferred input positions and spiral charges; black crosses mark the
    true input positions.
    Error annotations report the maximum relative spiral-charge error and, for the
    unknown-position case, the maximum position error normalized by the admissible
    disk radius \(R\).
    The simulations use a square domain of side length \(500\), with a
    central disk of radius \(100\), both in units of \(\ell/\sqrt{2}\).
 }
\label{fig:fig4}
\end{figure*}

Having validated the relative phase model in the two-defect setting, we next
apply it to a many-defect configuration. The linearity of
Eq.~\eqref{eq:phase-diffusion} makes it simple to design initial conditions
that produce prescribed transient textures. In the example in
Fig.~\ref{fig:fig3}, the initial \(\varphi\) combines a spiral-charge
component, corresponding to rotated defect orientations, with a localized
phase perturbation. The localized perturbation decays rapidly, while the
longer-wavelength content persists to intermediate times, producing a
frown-like texture that evolves through a smile-like state.  The reconstruction
agrees with the corresponding LdG relaxation, with small pointwise phase error
over the computed trajectory, showing that the~\mbox{relative} phase diffusion model
captures many-defect texture evolution beyond the two-defect geometry.
Importantly, such nontrivial time evolution cannot be explained within a quasistatic
approximation.

\begingroup
\setlength{\parskip}{0pt}
The linearity of Eq.~\eqref{eq:phase-diffusion} also makes inverse problems
tractable. We consider configurations in which some defects are designated as
hidden inputs and others as measured outputs. The input defects, located inside
the circles shown in Fig.~\ref{fig:fig4}(a),(d), are initially rotated by
unknown amounts, after which the system is allowed to relax. During this
relaxation, only noisy orientation measurements of the output defects are
available.
Synthetic measurements are generated from full nonlinear LdG dynamics, but the
inference is performed using the diffusion equation for the relative phase.
When the input positions are known, the unknown spiral
charges, and hence the initial input orientations, are recovered by linear least
squares from the output orientation traces
(Fig.~\ref{fig:fig4}(a)--(c)). When the input positions are unknown, we solve a
sparse recovery problem over a grid of candidate locations to infer the number
of inputs and estimate their positions and spiral charges. The nonzero
candidates are then refined off grid (Fig.~\ref{fig:fig4}(d)--(f)).
In both cases, the recovered initial relative phase agrees with the hidden state used to
generate the LdG data.
This demonstrates that unobserved defect configurations can be inferred from sparse
output measurements, suggesting a route to nonlocal readout of nematic-bit
states~\cite{kos2022nematic}.
\par
\endgroup

Our results show that the orientations of pinned nematic defects do not form an
autonomous quasistatic dynamical system. Instead, their relaxation is captured
by a bulk relative phase field whose linear diffusion reproduces both the
rotation of the pinned cores and the surrounding texture evolution.
This formulation provides a tractable linear model for defect-orientation
dynamics, enabling transient pattern design and the recovery of hidden defect
configurations from sparse measurements. Thus, defect orientations can serve as
controllable degrees of freedom in programmable nematic media, but their
dynamics are described by diffusion of the relative phase rather than by a
quasistatic torque law.

\begin{acknowledgments}
    We thank \v{Z}iga Kos for helpful discussions at an early stage of this work. This research received support through Schmidt Sciences, LLC (J.D.), and the MathWorks Professorship Fund (J.D.).
\end{acknowledgments}

\bibliographystyle{apsrev4-2}
\bibliography{references}

\end{document}